\documentclass[12pt]{article}
\usepackage{amsmath}
\usepackage{cite}
\usepackage{epsfig}
\usepackage{color}

\def\Journal#1#2#3#4{#1{\bf #2} (#4) #3}
\def\NPB{{\em Nucl.\ Phys.} B}

\def\PLB{{\em Phys.\ Lett.}  B}

\def\PRD{{\em Phys.\ Rev.} D}

\def\JHEP{{\em JHEP }}

\def\epj{{\em Eur.\ Phys.\ J.} C}

\def\PREP{{\em Phys.\ Rep. }}

\def\cO#1{{\cal{O}}\left(#1\right)}

\def\as{\alpha_{\mbox{\scriptsize s}}}

\def\be{\beta}

\def\out{\mbox{\scriptsize out}}
\def\PT{\mbox{\scriptsize PT}}\def\NP{\mbox{\scriptsize NP}}
\def\Ko{K_{\out}}
\def\qq{q\bar{q}}
\def\ee{e^+e^-}
\def\half{\mbox{\small $\frac{1}{2}$}}

\def\cD{{\cal{D}}}

\def\conf{\delta}
\def\cp{\lambda^{\NP}}

\def\cS{{\cal{S}}}

\def\VEV#1{\left\langle#1\right\rangle}

\def\re#1{(\ref{#1})}

\numberwithin{equation}{section}

 \newskip\humongous \humongous=0pt plus 1000pt minus 1000pt
   \newif\ifdtup

\def\ga{\mathrel{\mathpalette\fun >}}
\def\fun#1#2{\lower3.6pt\vbox{\baselineskip0pt\lineskip.9pt
  \ialign{$\mathsurround=0pt#1\hfil##\hfil$\crcr#2\crcr\sim\crcr}}}

\begin{document}
\begin{titlepage}
\renewcommand{\thefootnote}{\fnsymbol{footnote}}
\begin{flushright}
     Bicocca--FT-00/16 \\
     LPT-Orsay-00/88\\
     Pavia--FNT/T-00/18\\
     hep-ph/0010267 \\
     October 2000 \\
\end{flushright}
\par \vskip 10mm
\begin{center}
  {\Large \bf Near-to-planar 3-jet events \\[1mm]
in and beyond QCD perturbation theory}
\end{center}
\par \vskip 2mm
\begin{center}
{\bf 
A. Banfi$^{(1)}$, Yu.L. \ Dokshitzer$^{(2)}$\footnote{
On leave from St. Petersburg Nuclear Institute,
Gatchina, St. Petersburg 188350, Russia}, G.\ Marchesini$^{(1)}$ and
G. Zanderighi$^{(3)}$}\\[4mm]
\begin{minipage}{6.5in}
$^{(1)}$ Dipartimento di Fisica, Universit\`a di Milano--Bicocca
and INFN, Sezione di Milano, Italy \\[2mm]
$^{(2)}$ LPT, Universit\'e Paris Sud, Orsay, France \\[2mm]
$^{(3)}$ Dipartimento di Fisica, Universit\`a di Pavia
and INFN, Sezione di Pavia, Italy 
\end{minipage}
\end{center}

\par \vskip 2mm
\begin{center} {\large \bf Abstract} \\

\end{center}
\begin{quote}
  We present the results of QCD analysis of out-of-event-plane
  momentum distribution in 3-jet $\ee$ annihilation events. We
  consider the all-order resummed perturbative prediction and the leading
  power suppressed non-perturbative corrections to the mean value
  $\VEV{\Ko}$ and the distribution and explain their non-trivial
  colour structure.  
  The technique we develop aims at improving the accuracy of the
  theoretical description of multi-jet ensembles, in particular in
  hadron-hadron collisions. 
\end{quote}
\end{titlepage}

\section{Introduction \label{sec:Int}}
Physics of $\ee$ annihilation into hadrons is being used as a testing
ground for developing QCD analyses of multihadron production in hard
processes. The studies of the structure of hadron jets produced in
$\ee$ annihilation have led to establishing the standards for the
accuracy of QCD calculations.  The state-of-the-art includes
sophisticated perturbative (PT) analysis and taking into account the
leading non-perturbative (NP) effects. The latter show up as
contributions suppressed as a power of the hardness scale $Q$.  In
particular, in the distributions and means of various jet shapes the
NP effects are typically $\cO{1/Q}$.

At the level of the PT description
of $\ee$ jet-shape observables, 
these standards
include~\cite{PTstandards}
\begin{itemize}
\item
  all-order resummation of double- (DL) and single-logarithmic (SL)
  contributions due to soft and collinear gluon radiation effects,
\item 
  two-loop analysis of the basic gluon radiation probability and
\item 
  matching the resummed logarithmic expressions with the exact
  $\cO{\as^2}$ results.
\end{itemize}
The NP technology developed in recent years\cite{NPstandards}
unambiguously provides the {\em exponent}\/ of the leading power
correction to a given observable.  To quantify, in a universal way,
the {\em magnitude}\/ of the genuine confinement contribution one has
to
\begin{itemize}
  \item carefully address the problem of merging PT and NP
    contributions, and
  \item take into consideration two-loop effects in the
    magnitude of the NP correction, in particular those due to
    non-inclusiveness of jet observables (Milan factor).  
\end{itemize}
These are the today standards for the QCD predictions concerning
typical $\ee$ hadron systems, that is two-jet events. 

QCD treatment of hard processes involving hadrons in the {\em
  initial}\/ state such as DIS and, especially, hadron-hadron
collisions, seldom (if at all) reaches such an accuracy.
At the same time, hadron physics generated by ensembles of more than
two hard partons is theoretically quite rich.  Because of QCD
string/drag effects, the structure of accompanying flows of relatively
soft hadrons depends on the geometry and colour topology of the
multi-prong hard ``parton antenna''.  This makes it also practically
important:
measuring spatial distribution of hadron flows provides a
tool for identifying the nature of the underlying hard collision on
event-by-event basis \cite{event-by-event}, markedly at LHC.

NP effects should also be sensitive to the geometry and colour
structure of the multi-prong hard-parton system.  This issue, as far
as we can tell, has not yet been addressed in the literature.  As a
first step in this direction, we report in this letter the results of
the QCD analysis of the distribution of three-jet $\ee$ annihilation
events in out-of-event-plane transverse momentum $\Ko$.

We consider the near-to-planar ``3-jet region''
\begin{equation}
  \label{eq:TTM}
  T\sim T_M \gg T_m=\frac{\Ko}{Q}\>,
\end{equation}
where $T,T_M$, and $T_m$ are the thrust, thrust major and thrust
minor, respectively.  The thrust axis and the thrust-major axis are
set equal to the $z$- and $y$-axis respectively. The $\{yz\}$-plane is
defined as ``the event plane''.

At parton level the events in the region \eqref{eq:TTM} can be treated
as being generated by a system of energetic quark, antiquark and a
gluon accompanied by an ensemble of secondary (soft) partons.

We study the distribution of events in the cumulative out-of-plane
transverse momentum 
\begin{equation}
  \label{eq:Ko}
  \Ko=\sum_{a=1}^3|p_{ax}| + \sum_{i=1}^n |k_{ix}| ,
\end{equation}
with $p_1,p_2,p_3$ the momenta of the hard partons generating three
jets, and $k_i$ the secondary parton momenta.

Similar to 2-jet configurations in the region of two narrow jets
($1-T, M_H^2/Q^2,C,B_{T,W}\ll1$), physics of near-to-planar 3-jet
events, in the region $T_m\ll1$, involves DL (Sudakov) multiple
radiation effects.

Soft gluon radiation dominance makes it possible to resum DL terms to
all orders. The PT answer for the $\Ko$-distribution is expressed in
terms of (the exponent of) the ``radiators'' --- the basic
probabilities of a soft gluon emission off each of the three
event-generating hard partons.
Subleading SL corrections due to hard collinear radiation in one of
the jets can be embodied into the hard scale of the corresponding radiator.
SL effects due to soft inter-jet particle production induce
the dependence on the event geometry (angles between jets), both in
the PT and NP contributions.

\begin{figure}[ht]
  \begin{center}
    \epsfig{file=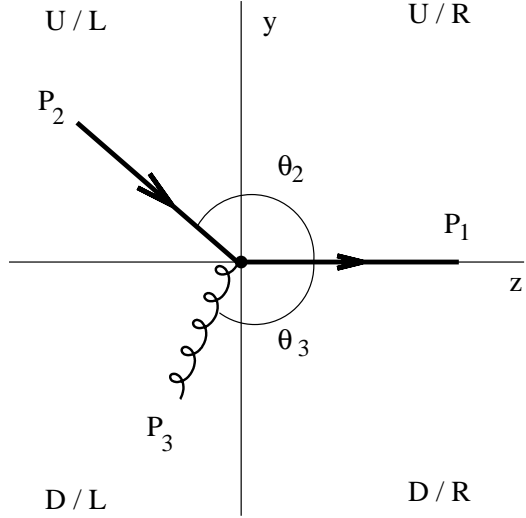,width=0.4\textwidth}
   \caption{The Born  configuration $\conf=3$ for $T=0.75$ and 
     $T_M=0.48$.  The thrust $T$ and thrust major $T_M$ are along the
     $z$- and $y$-axis respectively. The up, down, left and right
     hemispheres ($U,D,L,R$ respectively) are indicated.
\label{fig:conf3}}
  \end{center}
\end{figure}

At Born level there are only three massless partons with momenta
$P_a$ ($P_1>P_2>P_3$), so that $\Ko=0$.  The most energetic $P_1$
lies along the thrust axis, and we define the second most energetic
momentum $P_2$ to have a positive $y$-component.  Kinematics (energies
and relative angles) of three massless Born momenta is uniquely
determined by the $T$ and $T_M$ values.
There are three essentially different kinematical configurations of
the Born system, which we denote by $\conf$ ($\conf=1,2,3$), when $P_{\conf}$
is the gluon momentum.
In what follows we concentrate\footnote{The full answer
  will be given by the sum over the three configurations weighted by the
  relative 3-jet cross section (unless the gluon jet is tagged).  }
on the most probable configuration with
$\conf=3$ shown in Fig.~\ref{fig:conf3}, in which the gluon belongs to
the least energetic jet with momentum $P_3$.

The definition of the event plane beyond Born level involves momenta 
of all final particles, and the three hard partons with momentum
$p_a$, generally speaking, no longer lie in the plane: 
\begin{equation}
  \label{eq:recoil}
  p_a = P_a + q_a\>, \qquad \Ko=\sum_{a=1}^3 |q_{ax}|+ \sum_{i=1}^n |k_{ix}| ,
\end{equation}
with $q_{a}$ the recoil momenta of the hard partons.  Small
$x$-components of the recoil momenta enter the value of the observable
$\Ko$.
The {\em longitudinal}\/ (in-plane) components $q_{az}$, $q_{ay}$ are 
not necessarily small due to hard collinear jet splittings which make the 
final momentum $p_a$ a fraction of the initial Born $P_a$. As demonstrated in
\cite{acopt}, this rescaling of the longitudinal parton momenta
gets absorbed into the first hard correction to the emission
probability of soft gluons, which is then resummed and embodied into
the radiator.
Bearing this in mind, the recoil momenta can be treated as small.

The standard definition of the event plane as the plane formed by the
thrust and thrust-major axes leads to the 
four constraints involving the $x$- and $y$-components of particle
momenta:
\begin{equation}
  \label{eq:plane}
  \begin{split}
 &{q}_{1x}+\sum_R {k}_{ix}=0\>,\quad {q}_{1y}+\sum_R {k}_{iy}=0\>, \\
 &q_{2x}+q_{1x}\cdot \vartheta( q_{1y}) +\sum_U k_{ix}=0\>,\quad
 q_{3x}+q_{1x}\cdot\vartheta(-q_{1y}) +\sum_D k_{ix}=0\>.
\end{split}
\end{equation}
The definition of the right- ($R$), up- ($U$) and down- ($D$) hemispheres
is shown in Fig.~\ref{fig:conf3}.
Since all partons contribute to $\Ko$ and participate in the
kinematical relations, to resum and exponentiate secondary gluon
radiation one should resolve the additive constraints \re{eq:recoil}
and \re{eq:plane} with the help of Mellin and Fourier representations.

Given the complexity of the event plane conditions~\re{eq:plane}, the
analytic derivation of the resulting distributions turns out to be
technically rather involved, especially at the SL accuracy which is
needed to make quantitative predictions.
The essential features of the result, however, are rather interesting
and transparent from the point of view of the physics involved.  
In this letter we shall present the results and 
discuss their physical origin and properties. 
The detailed analysis to SL accuracy can be found
in~\cite{acopt,aconptobe}.

We calculate the {\em integrated}\/ cross section for fixed $T$,
$T_M\sim T$ and $T_m$ smaller than a given $\Ko/Q$.  
To SL accuracy, this cross section can be factorized as
\begin{equation}
  \begin{split}
\label{eq:Sig}
\frac{d\sigma(\Ko)}{dTdT_{M}} =(1+\cO{\as})
\cdot \sum_{\conf=1}^3 \frac{d\sigma_{\conf}^{(0)}}{dT dT_M} 
\cdot \Sigma_{\conf}(\Ko)\>,
\end{split}
\end{equation}
where ${d\sigma_{\conf}^{(0)}}/{dT dT_M}$ is the differential Born
3-jet cross section in the parton configuration $\conf$.  The factor
$\Sigma_{\conf}$ accounts for the soft radiation emitted by the hard
$\qq g$ system. 
The first factor is the (non-logarithmic) coefficient function.

In Section~2 we summarize the results of the PT analysis of the
$\Ko$-distribution obtained in~\cite{acopt}.
Section~3 is devoted to the leading $1/Q$ NP corrections to the
distribution and the mean. 
The logarithmic enhancement in the NP radiation depends on the
aplanarity of each of the hard partons ($\qq g$). These aplanarities,
in turn, are the result of the recoil against the PT radiation. Since
the structure of the recoil depends on the direction of the PT gluon,
one finds a peculiar colour structure of the NP corrections.
We give a simple physical explanation of this structure.
We conclude in Section~4.

\section{Perturbative $\Ko$-distribution \label{sec:PT}}

At the DL level, the integrated distribution $\Sigma$ in \re{eq:Sig}
is given simply by the Sudakov exponent
\begin{equation}
  \label{eq:CT}
  \Sigma^{\PT} (\Ko) \>\simeq \> 
\exp\left\{ -C_T \frac{\as}{\pi} \ln^2\frac{Q}{\Ko} \right\}, \quad
C_T\equiv \sum_{a=1}^3 C_a = 2C_F+N_c\,,
\end{equation}
where $C_a$ is the colour factor of parton $\#a$ in the given 
Born configuration.

With account of subleading effects the coupling starts to run, the
scales of individual parton contributions acquire different hard
correction factors and become geometry-dependent.  The recoil of hard
partons should be taken into account in the observable $\Ko$ and in the
event plane kinematics but, as shown in \cite{acopt}, can be neglected
at the PT level
in the soft radiation matrix elements.  The result at SL accuracy
reads
\begin{equation}
  \label{eq:SigPT}
\begin{split}
  \Sigma^{\PT} (\Ko) \>\simeq \>e^{-\sum_a
    R_a\left(\Ko^{-1}\right)}\cdot \cS\left(\as\ln
    \frac{Q}{\Ko}\right).
\end{split}
\end{equation}
The first factor is the exponent of the three two-loop parton
radiators $R_a$,
\begin{equation}
  \label{eq:Rdef}
  R_a(\Ko^{-1}) = C_a\, r\left(\Ko^{-1}, Q_a^{\PT}\right), \qquad 
  r(\Ko^{-1},Q_a^{\PT})=\frac2{\pi}
 \int_{\Ko}^Q\frac{dk_x}{k_x}\, {\as(2k_x)}\,
\ln \frac{Q_a^{\PT}}{k_x}\>,
\end{equation}
where the running coupling corresponds to the physical
(bremsstrahlung; CMW) scheme~\cite{scheme}.

In the configuration $\conf=3$ of Fig.~\ref{fig:conf3} we have
$C_1=C_2=C_F$, $C_3=N_c$, and the geometry-dependent PT scales
$Q_a^{\PT}$ are
\begin{equation}
  \label{eq:scalesPT}
  \left(Q_1^{\PT}\right)^2=\left(Q_2^{\PT}\right)^2 
= \frac{(P_1P_2)}{2}e^{-3/2}, \quad
  \left(Q_3^{\PT}\right)^2
 = \frac{(P_1P_3)(P_3P_2)}{2\,(P_1P_2)} e^{-\be_0/2N_c}. 
\end{equation}
The exponential factors here account for the SL hard corrections due
to collinear quark ($-3C_F$) and gluon ($-\be_0$) splittings.
Let us note that the characteristic gluon scale $Q_3$ is proportional
to the invariant gluon transverse momentum with respect to the $\qq$
pair. Notice that when the gluon \#3 becomes collinear to the quark
(antiquark) \#2, the scale $Q_3^{\PT}$ decreases, and the non-Abelian
contribution reduces.

\begin{figure}[ht]
  \begin{center}
    \epsfig{file=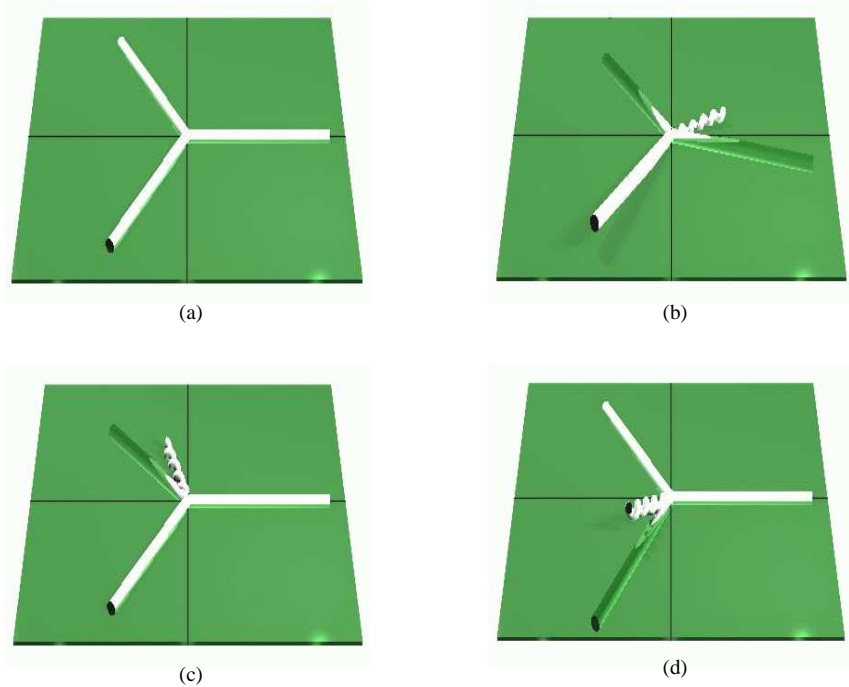,width=0.68\textwidth} 
   \caption{
     (a) Hard partons in a generic Born
     configuration.
     (b) Soft gluon $k$ (the short curly
     stick) is emitted in the right hemisphere. All three hard partons
     experience equal out-of-plane recoil, see \eqref{2.8}.  White
     (shadowed) partons have positive (negative) $x$-components of the
     momentum.
     (c) Soft gluon $k$ is emitted in the
     up-left region. According to \eqref{2.9}, only parton \#2 recoils
     (shadowed; has a negative $x$-components of the
     momentum).  (d) The case of $k$ in the down-left region.
\label{fig:recoil}}
  \end{center}
\end{figure}

The function $\cS$ in \re{eq:SigPT} is a subleading correction factor.
It depends on $\Ko$ via the SL function 
\begin{equation}
  \label{eq:r'}
  r'(\Ko^{-1}) \equiv \frac{d}{d\ln \Ko^{-1}} \, r(\Ko^{-1}) \simeq 
  \frac{2\as(\Ko)}{\pi}\ln\frac{Q}{\Ko}.
\end{equation}
To the first order in $r'$ one has \cite{acopt}
\begin{equation}
  \label{eq:cS}
  \cS = 1-\ln 2 \left(2C_1+C_2+C_3\right)
  \cdot r'(\Ko^{-1}) + \cO{{r}^{\prime 2}} ; \quad 
\left(2C_1+C_2+C_3\right)_{\conf=3}= 3C_F+N_c \,. 
\end{equation}
The fact that the contribution of the R-hemisphere jet \#1 is twice
larger than that of each of the L-hemisphere jets \#2,3 has a simple
kinematical origin.  The differential $\Ko$ distribution at first
order in $\as$ is determined by radiation of a single gluon with a
small transverse momentum $k_x$ off the $\qq g$ system. 
\begin{equation}
\frac{d\Sigma}{d\ln\Ko} = 
\frac{2\as}{\pi}\left\{\sum_{a}C_a\ln\frac{Q_a^{PT}}{\Ko}
+ (2C_1+C_2+C_3)\ln 2\right\}+\cO{\as^2} ,  
\end{equation}
where the first term originates from the DL radiator, and the second
one from the derivative of the SL factor $\cS$ in \re{eq:cS}.
So that the distribution can be written as 
\begin{equation}
\label{eq:first_ord}
\frac{d\Sigma}{d\ln\Ko} = 
\frac{2\as}{\pi}\left\{C_1\ln\frac{Q_1^{PT}}{\Ko/4} +  
C_2\ln\frac{Q_2^{PT}}{\Ko/2} + 
C_3\ln\frac{Q_3^{PT}}{\Ko/2} \right\}+\cO{\as^2} , 
\end{equation}
in full agreement with the kinematical conditions \re{eq:plane}.
Indeed, when the secondary gluon is emitted in the {\em right}\/
hemisphere (see Fig.~\ref{fig:recoil}b), all three hard partons
experience equal recoils, 
\begin{equation}\label{2.8}
\left.
\begin{split}
k_z>0, \quad & k_y>0: \quad   q_{1x}=-k_x=q_{2x}=-q_{3x}\quad\\
       & k_y<0: \quad   q_{1x}=-k_x=-q_{2x}=q_{3x}\quad
\end{split}
\right\}\> \Longrightarrow \> \Ko=4\cdot |k_x|\,.
\end{equation}
On the other hand, for the secondary gluon in the {\em left}\/
hemisphere (see Fig.~\ref{fig:recoil}c or d), only one hard parton
recoils against it:
\begin{equation}\label{2.9}
\left. 
\begin{split}
k_z<0, \quad &  k_y>0: \quad   q_{2x}=-k_x; \>\>  q_{1x}=q_{3x}=0\quad \\
             &  k_y<0: \quad   q_{3x}=-k_x; \>\> q_{1x}=q_{2x}=0\quad
\end{split}
\right\}\> \Longrightarrow \> \Ko=2\cdot |k_x|\,.
\end{equation}

As we shall see in the next section, similar kinematical effects show up
in the structure of the NP contribution as well.

\eqref{eq:first_ord} is an example of the rich information on the colour
structure and on the geometry of the underlying event provided by this
observable. 

\section{NP-effects in the distribution and mean \label{sec:NP}}

Similarly to other jet observables, the leading NP correction to the 
$\Ko$ distribution originates from radiation of small transverse
momentum gluons (``gluers'') by the $\qq g$ antenna. 
It is important that, at the NP level, the parton recoils $q_a$ not only
contribute to the observable (\eqref{eq:recoil}, \eqref{eq:plane}), 
as in the PT case, but also affect the radiator itself. 
In the linear approximation  in $1/Q$, the radiator acquires a NP
correction in the form
\begin{equation}
  \label{eq:RNP}
  R^{\NP} \>=\> \nu\cdot\cp\,\sum_{a=1}^3 C_a\,
  \ln\frac{Q^{\NP}_a}{|q_{ax}|}. 
\end{equation}
Here the variable $\nu$ in \re{eq:RNP} is Mellin conjugated to $\Ko$,
and $\cp$ (of dimension of mass) is a standard NP parameter which can
be related to the momentum integral of the QCD coupling in the
infrared region and parameterizes the $1/Q$ power correction.
$Q^{NP}_a\sim Q$ are the relevant scales which have the same geometric
structure as the PT scales \re{eq:scalesPT}, and differ from those by
a finite numerical factor~\cite{aconptobe}.

The logarithmic enhancement of the NP contribution \re{eq:RNP} is
similar to that for the jet Broadening observables (see~\cite{broad}).
It originates from the logarithmic integration over the gluer angle
$\theta$ with respect to the direction of the emitting parton $p_a$,
starting from large angles $\theta=\cO{1}$ down to the angle of the
hard parton $\theta_{\min}\sim |q_{ax}|/Q$. Radiation at smaller angles
corresponds to collinear splitting and does not contribute to the
observable due to real-virtual cancellation.
   
It is clear from the structure of the integral involving the Mellin
exponent $e^{\nu\Ko}$ and the exponent of the sum of the PT and NP
radiators that the correction term \re{eq:RNP} 
being linear in $\nu$ results in a {\em shift}\/ of the argument $\Ko$
of the PT distribution.  To find this shift one has to perform the
$q_{ax}$ integrals, for a given $\Ko$, taking into consideration the
phase space constraints
\re{eq:plane}.

To explain the structure of the answer let us introduce the
distributions $\cD_a(|q_{ax}|)$ of the parton recoil momenta. For the
leading jet \#1 it is given by the derivative of the DL Sudakov
exponent:
\begin{equation}
\label{eq:D1}
  \cD_1(q) = \frac{d}{d\ln q}\> e^{-R_1(q^{-1})}, \quad q=|q_{1x}|\,.
\end{equation}
The average value of the logarithm in \re{eq:RNP} (the logarithm of
the inverse quark angle) evaluated under the restriction
$|q_{1x}|<\Ko$ becomes
\begin{equation}
\label{eq:E1}
\begin{split}
\VEV{\ln\frac{Q_1^{\NP}}{|q_{1x}|}}_{|q_{1x}|<\Ko}
&\equiv e^{R_1(\Ko^{-1})}\int_0^{\Ko} \frac{dq}{q}\,
\ln\frac{Q_1^{\NP}}{q} \cD_1(q)
=\ln\frac{Q_1^{\NP}}{\Ko}+E_1(\Ko^{-1})\>,\\
E_1(\Ko^{-1})&\equiv 
\int_0^{\Ko} \frac{dq}{q}\, e^{-R_1(q^{-1})+R_1(\Ko^{-1})}.
\end{split}
\end{equation}
For the partons $\#2$ and $\#3$ in the $L$-hemisphere the situation is
more delicate.  As we already know from the plane constraints, in
order to inhibit the recoil of a left-hemisphere parton $\#a$ we have
to veto radiation both from $\#a$ and 
the right-hemisphere parton $\#1$.
Therefore,
\begin{equation}
\label{eq:Da}
\cD_a(q) = \frac{d}{d\ln q}\> e^{-R_a(q^{-1})-R_1(q^{-1})},
  \qquad a=2,3\,,
\end{equation}
and the average logarithms become
\begin{equation}
\begin{split}
\label{eq:Ea}
\VEV{\ln\frac{Q_a^{\NP}}{|q_{ax}|}}_{|q_{ax}|<\Ko}
&=\ln\frac{Q_a^{\NP}}{\Ko}+E_{a}(\Ko^{-1})\\
E_{a}(\Ko^{-1})&\equiv \int_0^{\Ko} \frac{dq}{q}\, 
e^{-R_a(q^{-1})-R_1(q^{-1})+R_a(\Ko^{-1})+R_1(\Ko^{-1})}\>, 
  \qquad a=2,3\,.
\end{split}
\end{equation}
In the limit of moderately small $\Ko$, so that
$\as\ln^2\frac{Q}{\Ko}\ll 1$, these functions can be approximately
evaluated to give
\begin{equation}
  \label{eq:Enum}
\begin{split}
E_1(\Ko^{-1})&=\frac{\pi}{2\sqrt{C_1\as(Q)}} -\ln\frac{Q}{\Ko}
\>+\>\cO{1}\,, \\
E_a(\Ko^{-1})&=\frac{\pi}{2\sqrt{(C_a+C_1)\as(Q)}} 
-\ln\frac{Q}{\Ko} \>+\>\cO{1}\,, \quad a=2,3\,.
\end{split}
\end{equation}
Starting from large values $\propto 1/\sqrt{\as}$, the functions $E$
decrease with $\Ko$.  They vanish as \mbox{$E\propto 1/r'(\Ko^{-1})$}
in the limit of extremely small $\Ko$, such that the Sudakov
suppression becomes very strong, $r'=
\frac{2\as}{\pi}\ln\frac{Q}{\Ko}\gg1$.  We conclude that the average
logarithms of the parton recoil angles in (3.1) stay at large constant
values $\cO{1/\sqrt{\as}}$, given by the the first terms on r.h.s.\ of
(3.6), in a broad region $1 \ga \Ko/Q \gg \exp(-1/\sqrt{\as})$, and
follow $\ln\frac{Q}{\Ko}$ for still smaller values of $\Ko$.  On the
basis of these considerations we can obtain the results for the
distribution and mean.

\subsection{NP correction to $\VEV{\Ko}$}
The NP correction to the mean value of $\Ko$ is given by the
$\nu$-derivative at $\nu=0$ of the NP radiator in \re{eq:RNP}.  To
evaluate $\VEV{\Ko}$ it suffices to take the unrestricted average of
$\ln(Q/|q_{ax}|)$ by setting $\Ko$ in \re{eq:E1}, \re{eq:Ea} at their
maximum values, i.e. $E_a(1/Q_a^{\NP})$.  The accurate answer that
includes first subleading corrections reads~\cite{aconptobe}
\begin{equation}
  \label{eq:mean}
  \VEV{\Ko} =  \VEV{\Ko}^{\PT}+ \cp \cdot\VEV{\Ko}^{\NP}\!\!\!, \quad
  \VEV{\Ko}^{\NP} = \sum_{a=1}^3 C_a\, E_a\left(\frac1{Q_a^{\NP}}\right)
  \left(1+\cO{\as}\right)\,.
\end{equation}
Analytically, 
\begin{equation}
\label{eq:VEV}
  \VEV{\Ko}^{\NP} \simeq 
 \frac{\pi}{2\sqrt{\as(Q)}} \left( \frac{C_1}{\sqrt{C_1}}
+ \frac{C_2}{\sqrt{C_1+C_2}} + \frac{C_3}{\sqrt{C_1+C_3}}\right)
\>+\> \cO{1}.
\end{equation}
Since combinations of charges appear in the denominators, we remark
that  the jets do not contribute independently to the mean at the NP level.

\subsection{NP shift of the PT ${\Ko}$-distribution}
The NP shift of the distribution
contains two structures, the additive and the non-additive pieces:
\begin{equation}
  \label{eq:shift}
  \Sigma(\Ko) = \Sigma^{\PT}(\Ko-\cp\delta K),  \quad 
  \delta K = \sum_{a=1}^3 C_a\left(\Delta_a+\Delta'_a\right).
\end{equation}
The additive contribution is
\begin{equation}
  \label{eq:d}
  \Delta_a(\Ko) = \ln\frac{Q_a^{\NP}}{\Ko} +\psi(1+R')+\gamma_E
  \>\simeq\> \ln\frac{Q}{\Ko}\,. 
\end{equation}
The other contribution has the following structure (a more accurate
expression can be found in~\cite{aconptobe}):
\begin{equation}
\begin{split}
  \label{eq:d'}
\Delta_1'(\Ko) &\simeq \frac{C_2+C_3}{C_T} \, E_1(\Ko^{-1}) \, , \\ 
\Delta_2'(\Ko) &\simeq \frac{C_3}{C_T} \, E_2(\Ko^{-1}) \, , \\ 
\Delta_3'(\Ko) &\simeq \frac{C_2}{C_T} \, E_3(\Ko^{-1}) \, .
\end{split}
\end{equation}
The additive contribution $\Delta$ has a simple origin. It corresponds
to the situation where all three emitters have recoil momenta of 
order $\Ko$. Therefore, this contribution to the shift follows
immediately from the NP radiator \re{eq:RNP} with $|q_{ax}|\sim \Ko$.
Such a 
situation is typical of well-developed parton systems with many
secondary partons.
In the kinematical region \mbox{$\as\ln^2\frac{Q}{\Ko}\gg1$}
the additive term $\Delta$ takes over $\Delta'$, and fully dominates
in the region $r'=\frac{2\as}{\pi}\ln\frac{Q}{\Ko}>1$:
\begin{equation}\label{additive}
 \delta K = \left(\, C_1+C_2+C_3\,\right)\ln\frac{Q}{\Ko}
\>+\>\cO{1}\,, \qquad \frac{\as}{\pi}\ln^2\frac{Q}{\Ko} \gg 1\,.
\end{equation}

On the contrary, for $\as\ln^2\frac{Q}{\Ko}\ll1$ there are few
secondary partons, and the non-additive contribution $\Delta'$ is
dominant.  To analyse this situation let us relate the NP correction
to the integrated spectrum $\Sigma$ to the {\em differential}\/
distribution
and expand the latter to first order in $\as$ 
in the region $\as\ln^2\frac{Q}{\Ko}\ll1$:
\begin{equation}
\Sigma-\Sigma^{\PT} \simeq -\cp \delta K \cdot \frac{d\Sigma^{\PT}}{d\Ko}
\simeq  -\cp \frac{R'}{\Ko} \cdot \delta K ; \quad R'=C_Tr'\,.
\end{equation}
Putting together the $\Delta$ and $\Delta'$ terms in $\delta K$ we can
then represent the answer as
\begin{equation}
  R'\cdot \delta K 
\simeq R'\cdot C_T\ln\frac{Q}{\Ko}
+ R_2'\cdot \left(C_1E_1+C_3E_3\right) 
+ R_3'\cdot \left(C_1E_1+C_2E_2\right), 
\end{equation}
where all the $E$-functions are evaluated at $\Ko^{-1}$. 
In the region under consideration we can use
the approximation \re{eq:Enum}, 
\begin{equation}
  E_a(\Ko^{-1}) = E_a(Q^{-1}) - \ln\frac{Q}{\Ko} 
\>=\> \VEV{\Ko}^{\NP}_a - \ln\frac{Q}{\Ko}, 
\end{equation}
where $\VEV{\Ko}^{\NP}_a$ is the NP contribution to the mean
associated with parton $\#a$ in \re{eq:mean}. 
Assembling the logarithmic pieces we arrive at
\begin{equation}
\begin{split}
 R'\cdot \delta K &\simeq 
  R_1'\cdot C_T \,\ln\frac{Q}{\Ko} \\
&+  R_2'\cdot\left( \VEV{\Ko}^{\NP}_1+ C_2\,\ln\frac{Q}{\Ko}
 + \VEV{\Ko}^{\NP}_3\right) \\
&+  R_3'\cdot\left( \VEV{\Ko}^{\NP}_1+ \VEV{\Ko}^{\NP}_2
+ C_3\,\ln\frac{Q}{\Ko}\right).
\end{split}
\end{equation}
Now we are ready to discuss the meaning of this expression.  

Making notice that the factor $R_a'$ corresponds to the radiation of a
perturbative gluon off the hard parton $\#a$, we see that the first
line describes PT emission in the $R$ hemisphere. In this case all
three hard partons experience equal recoil, $|q_{ax}|=\Ko/4$ (see
Fig.~\ref{fig:recoil}b), and the logarithmic factors in the NP
radiator \re{eq:RNP} are the same for all three gluers.

The second and third lines correspond to PT gluon emission off the
parton \#2 and \#3.  In the former case $|q_{2x}|=\Ko/2$, while the
other two partons remain in the plane ($q_{1x}=q_{3x}=0$) (see
Fig.~\ref{fig:recoil}c). Now, NP gluer radiation off \#2 brings in a
logarithmic correction as above.  At the same time, the logarithms
$\ln Q/|q_{ax}|$, $a=1,3$ are integrated with the corresponding
Sudakov distributions $\cD_a$ producing the contributions identical to
those to the mean.

Substituting $\VEV{\Ko}^{\NP}_a$ from \re{eq:VEV}, an approximate
expression for the NP shift becomes
\begin{equation}
\label{eq:SHIFT}
\begin{split}
 \delta K &= \left[\, \frac{(C_2+C_3)\sqrt{C_1}}{C_T}
 +\frac{C_2C_3}{C_T}\left(\frac{1}{\sqrt{C_1+C_2}}
 +\frac{1}{\sqrt{C_1+C_3}}\right) \right] \frac{\pi}{2\sqrt{\as(Q)}} \\
 &+\left[\,C_1+\frac{C_2^2+C_3^2}{C_T} \,\right]\ln\frac{Q}{\Ko}
\>+\>\cO{1}\,, \qquad \frac{\as}{\pi}\ln^2\frac{Q}{\Ko}\ll 1\,.
\end{split}
\end{equation}
The general expression for the NP shift
\re{eq:shift}--\re{eq:d'} interpolates between the two limiting regimes
\re{eq:SHIFT} and \re{additive}. 

\section{Conclusions \label{sec:Conc}}
We have presented the results and discussed the main physical
ingredients of the QCD analysis of the out-of-event-plane momentum
distribution in 3-jet events. It aims at improving the accuracy of the 
theoretical description of the final state structure of hard
processes involving more than two jets.  It should be possible to
generalize the present approach to analyse production of two
large-$p_\perp$ jets in DIS and production of a jet recoiling against
$W^\pm$, $Z^0$ or a large-$p_\perp$ photon in hadron-hadron collisions.

The most interesting feature of the $\Ko$ distribution is the
non-additivity of contributions due to the radiation off the three
jets.  In the PT contribution it occurs at the level of a subleading
SL correction, while in the $1/Q$ NP correction terms it shows up at
the leading level via peculiar combinations of colour charges like
$(C_F+C_A)^{-\half}$ (see \re{eq:VEV}, \re{eq:SHIFT}).
These can be traced back to the specific pattern of parton recoil 
due to the geometry of the event plane. 

Technical details of the calculations and the final expressions within
SL accuracy can be found in \cite{acopt,aconptobe}.  The relative
$\cO{\as}$ correction in \re{eq:Sig} has been computed \cite{agg} by
matching the expansion of the approximate resummed PT formula from
\cite{acopt} with the $\cO{\as^2}$ calculation of the differential
$\Ko$ distribution provided by four-parton event generators. (It is
possible, in principle, to further improve the result by including the
$\cO{\as^2}$ term in the non-logarithmic correction factor with the
help of five-parton generators.)

Predictions for the $\Ko$ distribution and means in $\qq\gamma$
events can be obtained from the above formulae by simply setting to
zero the colour charge corresponding to the gluon ($C_3=0$ in the
colour configuration of Fig.~1.)

Along the same lines the PT and NP predictions can be derived for the
mean and the distribution of $\Ko$ in the $R$-hemisphere, rather than
$\Ko$ of the event as a whole.
The corresponding perturbative formulae are much simpler, and so are
the NP expressions.  They can be essentially obtained by setting
$C_2=C_3=0$ in the above expressions for the total $\Ko$.  (There is an
additional difficulty, though, in calculating the proper scales
$Q^{\PT}$ and $Q^{\NP}$ with subleading accuracy, due to kinematical
complications.)

\vspace{-0.7cm}
\paragraph{Acknowledgements}
We are grateful to Gavin Salam for discussions and suggestions.

\end{document}